\documentclass[]{article}
\usepackage{proceed2e}
\usepackage{graphicx} % more modern
\usepackage{subfigure}
\usepackage{multirow}
\usepackage{amsmath}
\usepackage{hyperref}
\usepackage{algorithm}
\usepackage{algorithmic}
\usepackage{slashbox}
\title{Dynamic Infinite Mixed-Membership Stochastic Blockmodel}

\author{} % LEAVE BLANK FOR ORIGINAL SUBMISSION.
\begin{document}

\maketitle

\begin{abstract}
Directional and pairwise measurements are often used to model inter-relationships in a social network setting. The Mixed-Membership Stochastic Blockmodel (MMSB) was a seminal work in this area, and many of its capabilities were extended since then. In this paper, we propose the \emph{Dynamic Infinite Mixed-Membership stochastic blockModel (DIM3)}, a generalised framework that extends the existing work to a potentially infinite number of communities and mixture memberships for each of the network's nodes. This model is in a dynamic setting, where additional model parameters are introduced to reflect the degree of persistence between one's memberships at consecutive times. Accordingly, two effective posterior sampling strategies and their results are presented using both synthetic and real data.
\end{abstract}

\section{Introduction}
Community learning is an emerging topic applicable to many social networking problems, and has recently attracted research interest from the machine learning community.
Many models were proposed in the last few years. Some notable earlier examples include \emph{Stochastic BlockModel} \cite{nowicki2001estimation} and \emph{Infinite Relational Model} \cite{kemp2006learning} where they aim to partition a network of nodes into different groups based on their pairwise and directional binary observations.

To address the phenomenon that relationships between nodes may change over times, the recent work in this area focuses on the ``dynamic'' settings. For example, \cite{Yangml11} used the stochastic blockmodel to model the evolving community's behaviour over times. The work, however, assumes a fixed number of $K$ communities exist where a node $i$ can potentially belong to. However, in many applications, an accurate guess of $K$ can be impractical.

Infinite Relational Model was incorporated \cite{conf/nips/IshiguroIUT10} to address this problem, where $K$ can be inferred from the data itself. However, just as \cite{kemp2006learning}, its drawback is that the model assumes each node $i$ must belong to only a single community $k$ (i.e., $z_i=k$). Therefore, a relationship between nodes $i$ and $j$ can only be determined from their community indicators $z_i$ and $z_j$. This approach can be inflexible in many scenarios, such as the monastery example depicted in \cite{airoldi2008mixed}. To this end, the authors in \cite{airoldi2008mixed} introduced the concept of mixed-membership, where they assume each node $i$ may belong to multiple communities. The membership indicator is no longer sampled from each pair of community indicators $z_i$ and $z_j$. Instead, they are sampled from pairs of interactions between nodes $i$ and $j$. A few variants were subsequently proposed from MMSB, examples include: \cite{fu2009dynamic, xing2010state} extends the mixture-membership model with a dynamic setting; \cite{koutsourelakis2008finding} extends the MMSB into the infinite case; and \cite{kim2012nonparametric} incorporates the node's metadata information into MMSB.

In practice, the above discussed aspects (infinite, dynamic, mixture membership and data-driven inference) are often embedded into one complex network environment, as seen in the increasing social networking activities. However, there is no work reported on addressing all these aspects towards a flexible and generalised framework. To this end, we feel the emergent need to effectively unify these above mentioned models and to provide a flexible and generalised framework which can encapsulate the advantages from most of these works. Accordingly, we propose the \emph{Dynamic Infinite Mixed-Membership stochastic blockModel (DIM3)}. DIM3 allows the following features.
Firstly, it allows the infinite number of communities; secondly, it allows mixed-membership for each node; thirdly, the model extends to the dynamic settings. Lastly, it is apparent that in many social networking applications, a node's membership may become persistent over consecutive times, for example, a person's opinion of his peer is more likely to be consistent in two consecutive times.

To model persistence, we here devise two different implementations. The first is to have a single membership distribution for each node at different time intervals. The persistence factor is dependent on the statistics of each node's interactions with the rest of the nodes. The second implementation is to allow a set of mixed-membership distributions to associate with each node, and they are time-invariant. The number of elements in the set varies non-perimetrically similar to that of used in \cite{fox2008hdp}. The persistence factor is dependent on the value of membership indicator at the previous time.

%Therefore, we have deployed a stickiness parameter in our model, similar to that used in the sticky HDP-HMM's \cite{fox2008hdp}.

Two effective sampling algorithms are consequently designed for our proposed models, using either the Gibbs and Slice sampling technique for efficient model inference.

The rest of the article is organised in the following: Section \ref{sec_2} introduces our main framework and explains how it can incorporate infinite communities in a dynamic setting. The two models are explained, and their inference schemes are also detailed in Section \ref{sec_3}. In Section \ref{sec_4}, we show the experimental results of the several proposed models using both the synthetic and real-world social network data. Conclusions and future works can be found in Section \ref{sec_5}.

\section{The DIM3 Model} \label{sec_2}

\subsection{Notations}
For the notational clarity, we define all the symbols first, in which they will frequently appear in various sections in the rest of the paper. We use $E=\{e_{ij}^t\}_{n\times n}^{1:T}$ to denote the entire set of directional and binary observations: if $i$ has a relationship to node $j$ at time $t$, it implies $e_{ij}^t = 1$. Otherwise, $e_{ij}^t = 0$. Note that the directional relation $e_{ij}^t$ discussed here is specific to each pair of communities membership indicators $(s_{ij}^t , r_{ij}^t)$. For each pair of nodes, $i$ and $j$, at time $t$, $s_{ij}^t$ refers to the sender's community membership indicator. Correspondingly, $r_{ij}^t$ is for the receiver's community membership indicator. For the reason of simplicity and also making notations inline with what was used in the traditional HDP literature,  we use $Z$ to denote all the hidden labels $\{s_{ij}^t, r_{ij}^t\}$.

For each node $i$ at time $t$, there is a mixed-membership distribution, $\boldsymbol{\pi}_i^t$ having infinite components, and the $k^{\text{th}}$ component of $\boldsymbol{\pi}_i^t$, i.e., $\boldsymbol{\pi}_{ik}^t$ represents the ``significance'' of community $k$ for  node $i$.

There is also a role-compatibility matrix $W$ used. As the number of communities can become potentially infinite, the dimension of $W$ can potentially be $\infty \times \infty$ where its $(k,l)^{\text{th}}$  entry, i.e., $W_{k,l}$ represents compatibilities between communities $k$ and $l$.  Commonly, one assumes that each $W_{k,l}$ is i.i.d from $Beta(\lambda_1, \lambda_2)$  which gives conjugacy to the Bernoulli distribution used to generate $e_{ij}^t$ \cite{kim2012nonparametric}.

We use $n_{k,l}^t$ to denote the number of links from communities $k$ to $l$, i.e., the number of times in which $s^t_{ij} = k$ and $r^t_{ij} = l$ simultaneously. We let $n_{k,l}^t = n_{k,l}^{t,1} + n_{k,l}^{t,0}$. $n_{k,l}^{t,1}$ denotes the part of $n_{k,l}^t$ where the corresponding $e^t_{ij} = 1$. The number of times that a node $i$ has participated in community $k$ (both as a sending and receiving) at time $t$ is represented by $N_{ik}^t$.

\subsection{Mixture Time Variant (MTV) and Mixture Time Invariant (MTI) Models}

To address the phenomenon that one's social community's memberships may change over times, in our \emph{DIM3} model, we allow each node's mixed-membership indicators to change cross times. Additionally, it is imperative that these indicators should have some persistence with its past values which reflects the reality of social behaviour.

The modelling is achieved in two ways. The first is to allow the mixed-membership distributions itself to change over times. However, there is only a single (but different) distribution for each node at time $t$. The membership indicator of a node at time $t$ is dependent on the ``statistics'' of all membership indicators of the same node at $t-1$ and $t+1$. This is illustrated in the ``Mixture Time Variant (MTV)'' version.

The second method is to allow the mixed-membership distributions to stay invariant over times. However, there may be infinitely-possible many distributions associated with each node, but due to a HDP prior, often, only a few distributions will be discovered. This is illustrated in the ``Mixture Time Invariant (MTI)'' model. In this case, the membership indicator at time $t$ is dependent and more likely to have the same value as it was in $t-1$.

In both cases, the persistence effect is achieved through a sticky parameter $\kappa$ which is added to alter the membership distributions.

\subsection{Mixture Time Variant (MTV) Model} \label{sec_22}

%As described in the introduction part, we also intend to model the natural phenomenon that one's social community membership do not change abruptly between consecutive time intervals. Therefore, a sticky parameter $\kappa$ is introduced to incorporate the node's previous activities into the current mixed-membership distribution.
%
%This explicit modelling of time-based correlation gives us extra flexibility in comparison with the traditional models \cite{fu2009dynamic, xing2010state, DBLP:journals/jmlr/HoSX11}, where they handle time-correlation implicitly. The infinitely-possible communities is achieved by using a popular Hierarchical Dirichlet Process \cite{teh2006hierarchical} prior.

\begin{figure}[htbp]
\centering
\includegraphics[scale=0.45, width = 0.45 \textwidth, bb = 143 506 323 667, clip]{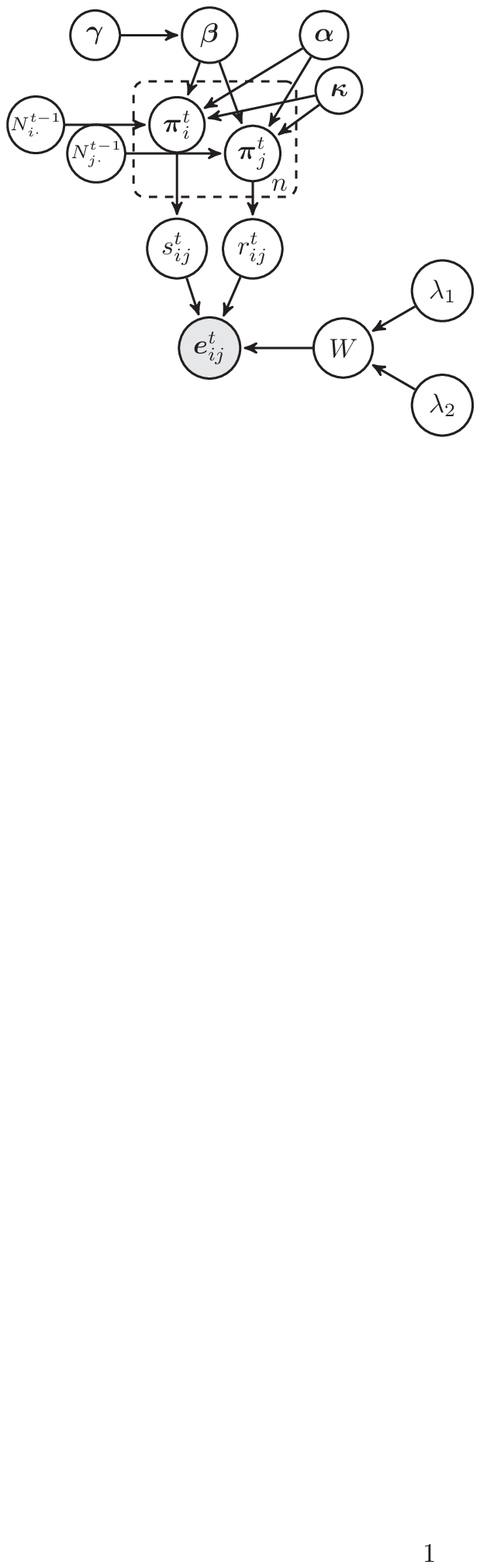}
\caption{The \emph{MTV-DIM3} Model}
\label{fig_5}
\end{figure}

In Figure \ref{fig_5}, we show the graphical model of the \emph{MTV-DIM3} model.  Here we only show all the variables involved for time $t$, and omit the other times, where the structure is identical.

The corresponding generative process is provided as follows:
\begin{enumerate}
  \item Global Setting - where its value is shared across all times $1:T$.
  \begin{itemize}
  \item $\boldsymbol{\beta}\sim GEM (\gamma)$
  \item $W_{k,l}\sim Beta(\lambda_1, \lambda_2) \; \forall k,l$
  \end{itemize}
  \item Mixed-membership distribution
  \begin{itemize}
  \item $\boldsymbol{\pi}_i^t\sim DP \left(\alpha+\kappa, \frac{\alpha\boldsymbol{\beta}+\frac{\kappa}{2n}\cdot\sum_kN_{ik}^{t-1}\boldsymbol{\delta}_k}{\alpha+\kappa} \right)$ denotes a node $i$'s mixed-membership distribution at time $t$.
  \end{itemize}
  \item Relationship Sampling
  \begin{itemize}
  \item For each pair of $i,j\in\{1, \cdots, n\}, t\in\{1, \cdots, T\}$
  \begin{itemize}
  \item $s_{ij}^t\sim Multi (\boldsymbol{\pi}_i^t)$: sending community's membership indicator;
  \item $r_{ij}^t\sim Multi (\boldsymbol{\pi}_j^t)$: receiving community's membership indicator;
  \item $e_{ij}^t\sim Bernoulli (W_{s_{ij}^t,r_{ij}^t})$ relation from nodes $i$ to $j$ at time $t$.
  \end{itemize}
  \end{itemize}
\end{enumerate}
Here $N_{ik}^{t-1}=\sum_{l=1}^N \boldsymbol{1}(s_{il}^{t-1}=k)+\sum_{l=1}^N \boldsymbol{1}(r_{li}^{t-1}=k)$, representing the count number that a node $i$ has been associated with a community $k$ at time $t-1$.

$\boldsymbol{\beta}$ is used as a global random variable, representing the ``significance'' of all existing communities at all times, while $W$ is the communities' compatibility matrix as described previously. As the prior $P(W)$ is element-wise $Beta$ distributed, which is conjugate to the Bernoulli distribution $P (e^t_{i,j} | .)$. Therefore, we can obtain a marginal distribution of $P (e^t_{i,j})$, i.e., $\int_{W} p (e^t_{i,j} | W) p(W) d(W)$ analytically, and hence do not need to explicitly sample values of $W$.

The mixed-membership distribution $\{\boldsymbol{\pi}_i^t\}_{1:n}^{1:T}$ is sampled from the Dirichlet Process with a concentration parameter $(\alpha+\kappa)$ and a base measure $\frac{\alpha\boldsymbol{\beta}+\frac{\kappa}{2n}\sum_kN_{ik}^{t-1}\boldsymbol{\delta}_k}{\alpha+\kappa}$.  There will be $N \times T$ of these distributions. They jointly describe each node's activities. It should be noted that each $\boldsymbol{\pi}_i^t$ is responsible to generate both the senders' label $\{s_{ij}^t\}_{j=1}^n$ from node $i$ and receivers' label $\{r_{ji}^t\}_{j=1}^n$ to node $i$.

In the base measure, the introduced sticky parameter $\kappa$ stands for each node's time influence on its mixed-membership distribution. In another words, we assume that each node's mixed-membership distribution at time $t$ will be largely influenced by its activities at time $t-1$. This is reflected in the hidden label's multinomial distribution that the previous explicit activities will occupy a fixed proportion $\frac{\kappa}{\alpha+\kappa}$ to the current distribution. The larger the value of $\kappa$, the more weight that the activities at $t-1$ is going to play at time $t$.

As our method is largely based on the HDP framework, therefore, we will use the popular ``Chinese Restaurant Franchise (CRF)'' \cite{teh2006hierarchical, fox2008hdp} analogy to further explain our model. Using the CRF analogy, the mixed-membership distribution associated with a node $i$ at time $t$ can be seen as a restaurant $\boldsymbol{\pi}_i^t$, with its dishes representing the communities. If a customer $s_{ij}^t( \text{or } r_{ji}^t)$ eats the dish $k$ at the $i^{\text{th}}$ restaurant at time $t$, then $s_{ij}^t(r_{ji}^t)=k$. $\forall t>1$, the restaurant $\boldsymbol{\pi}_i^t$ would have its own specials on the served dishes, representing the ``sticky'' configuration in the graphical model. Contrast to the sticky HDP-HMM \cite{fox2008hdp} approach, which places special on one dish only, in our work, we allow multiple specials, where the weight of each special dish is adjusted according to the number of served dishes at this restaurant at time $t-1$, i.e., $\frac{\kappa}{2n}\sum_kN_{ik}^{t-1}\boldsymbol{\delta}_k$. Therefore, we can ensure that the special dishes are served persistently across times in the same restaurant.

% are added in accordance with the dish choices of the previous time.
  %restaurant $\boldsymbol{\pi}_i^t$ could served the current customers $\{s_{ij}^t, r_{ji}^t\}_{j=1}^n$ while keep the restaurant's dish choices
%We did not consider the communities' relationship dynamic changing for simplicity since it is not our focus. We put a beta distribution on the role-compatibility matrix $W$. In this way, we integrate out the explicit $W$'s value and take the prior parameter directly into the calculation.

\subsection{Mixture Time Invariant (MTI) Model}
\begin{figure}[htbp]
\centering
\includegraphics[scale=0.45, width = 0.45 \textwidth, bb = 142 479 355 671, clip]{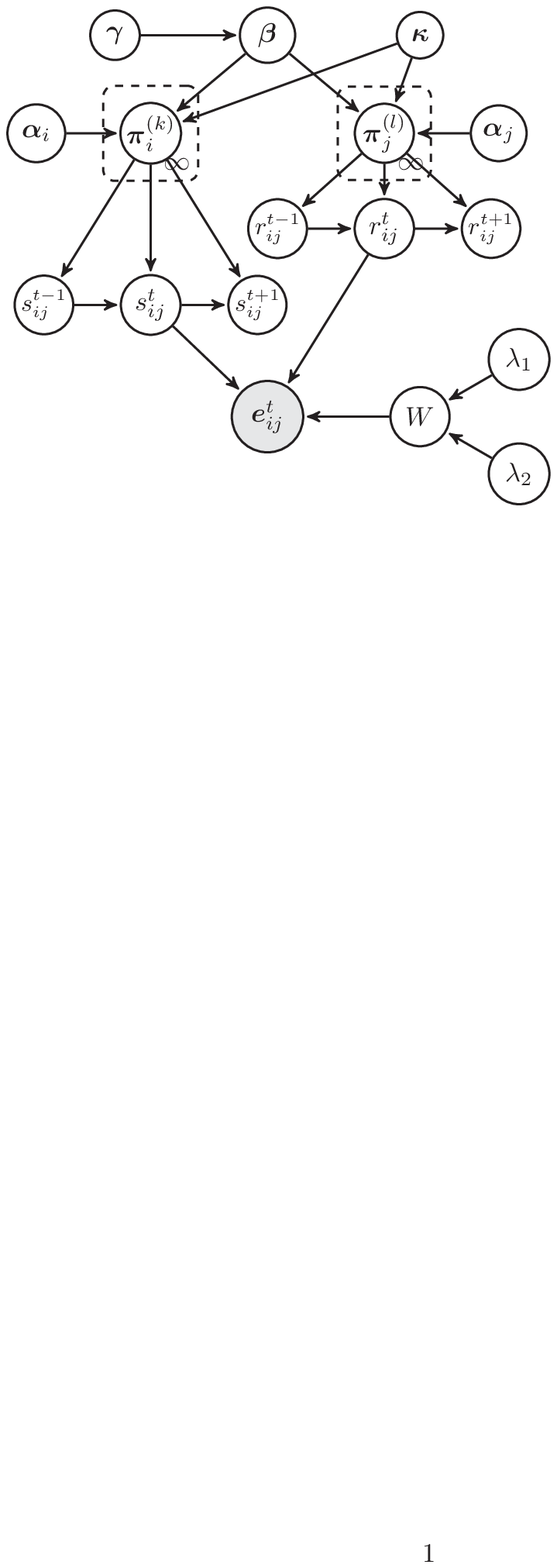}
\caption{The \emph{MTI-DIM3} Model}
\label{fig_3}
\end{figure}

%Another optional time-correlated relation depiction method is to use an HDP-HMM \cite{teh2006hierarchical} configuration on each of the node. While each $s_{ij}^t$ ($r_{ij}^t$) follows this time routine, their are indexed from a shared mixed-membership distribution $\boldsymbol{\pi}_i^{(s_{ij}^{t-1})}$ on node $i$.
We show the \emph{MTI-DIM3} model in Figure \ref{fig_3}.

In this model, each node has a variable number of membership distributions associated with it, which may potentially be infinite. At each time $t$, its membership indicator $s_{ij}^t$ is generated from ${\bf \pi}_{s_{ij}^{t-1}}$. In order to encourage persistence, each $\pi_{ik}$ was generated from a corresponding $\beta$, where $\kappa$ was added to $\beta$'s $k^{\text{th}}$ component \cite{fox2008hdp, fox2011bayesian, fox2011sticky}.

The corresponding generative process of the \emph{MTI-DIM3} model is provided as follows:
\begin{enumerate}
  \item Global Setting - where its value is shared across all times $1:T$.
  \begin{itemize}
  \item $\boldsymbol{\beta}\sim GEM (\gamma)$
  \item $W_{k,l}\sim Beta(\lambda_1, \lambda_2), \forall k,l$
  \end{itemize}
  \item Mixed-membership distribution
  \begin{itemize}
  \item $\boldsymbol{\pi}_i^{(k)}\sim DP \left(\alpha_i+\kappa, \frac{\alpha_i\boldsymbol{\beta}+\kappa\boldsymbol{\delta}_k}{\alpha_i+\kappa} \right)$ denotes a node $i$'s mixed-membership distribution.
  \end{itemize}
  \item Relationship Sampling
  \begin{itemize}
  \item For each pair of $i,j\in\{1, \cdots, n\}, t\in\{1, \cdots, T\}$
  \begin{itemize}
  \item $s_{ij}^t\sim Multi ({\boldsymbol{\pi}}_i^{(s_{ij}^{t-1})})$: sending community's membership indicator;
  \item $r_{ij}^t\sim Multi ({\boldsymbol{\pi}}_j^{(r_{ij}^{t-1})})$: receiving community's membership indicator;
  \item $e_{ij}^t\sim Bernoulli (W_{s_{ij}^t,r_{ij}^t})$ relation from nodes $i$ to $j$ at time $t$.
  \end{itemize}
  \end{itemize}
\end{enumerate}

$\boldsymbol{\beta}$ and $W$'s generation is the same as in Section \ref{sec_22}.
The set of membership indicators $\{s_{ij}^t, r_{ji}^t|j=1,\cdots,n,t=1,\cdots, T\}$ will be sampled from the time-invariant mixed-membership distribution set, $\{\boldsymbol{\pi}_i^{(k)}\}_{k=1}^{\infty}$, where each member is independently distributed from a Dirichlet Process with a concentration parameter $(\alpha+\kappa)$ and a base measure $\frac{\alpha\boldsymbol{\beta}+\kappa\boldsymbol{\delta}_k}{\alpha+\kappa}$.

At time $t$, a membership indicator $s_{ij}^t (\text{or } r_{ji}^t)$ is sampled from the distribution $\boldsymbol{\pi}_{i}^{(s_{ij}^{t-1})}(\text{or } \boldsymbol{\pi}_{i}^{(r_{ji}^{t-1})}) \forall i\in\{1,\cdots,n\}$.

Back to the CRF \cite{teh2006hierarchical} analogy, we have $N\times\infty$ matrix, where its $(i,k)^{\text{th}}$ element refers to $\pi_i^{(k)}$, which can be seen as the weights of eating each of the available dishes. A customer $s_{ij}^t (\text{or } r_{ji}^t)$ therefore can only travel between restaurants located at the $i^{\text{th}}$ row of the matrix. When $\pi_i^{(k)}$'s $k^{\text{th}}$ component is more likely to be larger, it means that the dish $k$ is a special dish for restaurant $k$. Therefore, a customer is at restaurant $k$ at time $t-1$, is more likely to eat the same dish ( i.e., $k^{\text{th}}$ dish), and hence to stay at restaurant $k$ at time $t$.

%We use the sticky HDP-HMM's configuration to complete the generation of the mixed-membership distribution $\{\boldsymbol{\pi}_i^k\}_{1:n}^{\infty}$. $\forall i,k$, each $\boldsymbol{\pi}_i^k$ is sampled from the Dirichlet Process with concentration parameter $\alpha+\kappa$ and base measure $\frac{\alpha\boldsymbol{\beta}+\kappa\boldsymbol{\delta}_k}{\alpha+\kappa}$. More specifically, we partition all of the senders' label $\{s_{ij}^t\}_{j=1:n}^{1:T}$ and receivers' label $\{r_{ij}^t\}_{j=1:n}^{1:T}$ equally into $n$ parts, according to their belonging node, and put a ``component'' shared sticky HDP-HMM, noted as $H_i, \forall i\in 1,\ldots, n$ on each part.
%

\section{Inference} \label{sec_3}
Two sampling schemes are implemented to complete the inference on \emph{MTV-DIM3}: the standard Gibbs sampling and Slice-Efficient sampling, which both target the posterior distribution. Due to the space limit, we do not present here the detailed sampling scheme of the \emph{MTI-DIM3}. Interesting readers can refer to the supplementary material . Due to the double-blind review policy, we anonymously put the supplementary material in \footnote{\href{https://dl.dropbox.com/u/56306996/icml2012stylefiles/sampling\%20scheme/supplementary\%20materials.pdf}{Here is the anonymous link address.}}
 %Only eliminating the two `$\backslash$' symbols in the URL address would be valid, sorry for this inconvenience.}.

%Our gibbs sampling's infinite state realization is similar to \cite{kim2012nonparametric}, in which they claimed they were using the retrospective MCMC way, however without involving the strict acceptance criterion in the common Retrospective MCMC. So we still call our method as the gibbs sampling.

\subsection{Gibbs Sampling}
The Gibbs Sampling scheme is largely based on \cite{teh2006hierarchical}). The variables of interest are: $\boldsymbol{\beta}$, $Z$ and auxiliary variables $\hat{\boldsymbol{m}}$, where $\hat{\boldsymbol{m}}$ refers to the number of tables eating dish $k$ as used in \cite{teh2006hierarchical, fox2008hdp} without counting the tables that generated from the sticky portion, i.e., $\kappa N^{t-1}_{ik}$. Note that we do not sample $\{\boldsymbol{\pi}_i^t\}_{1:n}^{1:T}$, as it gets integrated out.

\begin{description}
\item{Sampling $\boldsymbol{\beta}$}

$\boldsymbol{\beta}$ is the prior for all $\{\boldsymbol{\pi}_i^t\}s$, which can be thought as the ratios between the community components for all communities. Its posterior distribution is obtained through the auxiliary variable $\hat{\boldsymbol{m}}$:

\begin{equation}
(\boldsymbol{\beta}_1, \cdots, \boldsymbol{\beta}_K, \boldsymbol{\beta}_{\mu})\sim Dir(\hat{\boldsymbol{m}}_{\cdot1}, \cdots, \hat{\boldsymbol{m}}_{\cdot K}, \gamma)
\end{equation}

where its detail can be found in \cite{teh2006hierarchical}.

\item{Sampling $\{s_{ij}^t\}_{n\times n}^{1:T}$, $\{r_{ij}^t\}_{n\times n}^{1:T}$}

Each observation $e_{ij}^t$ is sampled from a fixed Bernoulli distribution, where the Bernoulli's parameter is contained within the role-compatibility matrix $W$ indexed (row and column) by a pair of corresponding membership indicators $\{s_{ij}^t, r_{ij}^t\}$. W.o.l.g, $\forall k,l\in\{1, \cdots, K+1\}$, the joint posterior probability of $( s_{ij}^t=k, r_{ij}^t=l)$ is:

\begin{equation} \label{eq_2}
\begin{split}
& P(s_{ij}^t=k, r_{ij}^t=l|{Z}\backslash\{s_{ij}^t, r_{ij}^t\}, e, \boldsymbol{\beta}, \alpha, \lambda_1, \lambda_2, \kappa)\\
 \propto & P(s_{ij}^{t}=k|\{s_{ij_0}^t\}_{j_0\neq j},\{r_{j_0i}^t\}_{j_0=1}^n,  \boldsymbol{\beta}, \alpha, \kappa, N_i^{t-1})\\
 \cdot &\prod_{l=1}^{2n}P(z_{il}^{t+1}|z_{i\cdot}^t\slash s_{ij}^t,s_{ij}^t=k, \boldsymbol{\beta}, \alpha, \kappa, N_i^{t+1})\\
 \cdot & P(r_{ij}^{t}=l|\{r_{i_0j}^t\}_{i_0\neq i}, \{s_{ji_0}\}_{i_0=1}^n, \boldsymbol{\beta}, \alpha, \kappa, N_j^{t-1})\\
 \cdot & \prod_{l=1}^{2n}P(z_{jl}^{t+1}|z_{j\cdot}^t\slash r_{ij}^t,r_{ij}^t=l, \boldsymbol{\beta}, \alpha, \kappa, N_j^{t+1})\\
 \cdot & P(e_{ij}^t|E\backslash\{e_{ij}^t\}, s_{ij}^t=k, r_{ij}^t=l, \boldsymbol{Z}\backslash\{s_{ij}^t, r_{ij}^t\}, \lambda_1, \lambda_2)\\
\end{split}
\end{equation}

Detailed derivations of Eq. (\ref{eq_2}) is found at the supplementary materials.

Assuming the current sample of $\{s_{ij}^t, r_{ij}^t\}$ having values ranging between $1 \dots K$, we let undiscovered (new) community to be indexed by $K+1$. Then, to sample a pair $(s_{ij}^t, r_{ij}^t)$ in question, we need to calculate all $(K+1)^2$ combinations of values for the pair.

\item{Sampling $\hat{\boldsymbol{m}}$}

Using the restaurant-table-dish analogy, we denote $\boldsymbol{m}_{ik}^t$ as the number of tables eating dish $k \; \forall i,k,t$. This is related to the variable $\hat{\boldsymbol{m}}$ used in sampling ${\boldsymbol{\beta}}$, but also including the counts of the ``un-sticky'' portion, i.e., $\alpha\boldsymbol{\beta}_k$.

The sampling of $\boldsymbol{m}_{ik}^t$ is to incorporate a similar strategy as \cite{teh2006hierarchical, fox2008hdp}, which is independently distributed from:

\begin{equation} \label{eq_3}
\begin{split}
& \Pr(\boldsymbol{m}_{ik}^t=m|\alpha, \boldsymbol{\beta}_k, N_{ik}^{t-1}, \kappa)\\
\propto & S(N_{ik}^t, m)(\alpha\boldsymbol{\beta}_k+\kappa N_{ik}^{t-1})^m
\end{split}
\end{equation}
Here $S(\cdot, \cdot)$ is the Stirling number of first kind.

%In our distribution-sticky model, the ``sub-partition'' on the $k$-th community is influenced by two factors:

For each node, the ratio of generating new tables can result from two factors: (1) Dirichlet prior with parameter $\{\alpha, \boldsymbol{\beta}\}$ and (2) the sticky configuration from membership indicators at $t-1$, i.e., $\kappa N_{ik}^{t-1}$.

To sample $\boldsymbol{\beta}$, we need to only include tables generated from the ``un-sticky'' portion, i.e., $\hat{\boldsymbol{m}}$, where each $\hat{\boldsymbol{m}}^t_{ik}$ can be obtained from a single Binomial draw:

%The counting number of tables,
%on the first factor is our target as it provides's likelihood, while the second must be eliminated from $\boldsymbol{m}$.
%For $t>1$, let $\omega_{ik}^t$ being the ``sub-partition'' counts in the second factor of sticky membership indicator %configuration, we then have

\begin{equation} \label{eq_4}
\hat{\boldsymbol{m}}_{ik}^t\sim Binomial (\boldsymbol{m}_{ik}^t, \frac{\alpha\boldsymbol{\beta}_k}{\frac{\kappa}{2n} N_{ik}^{t-1}+\alpha\boldsymbol{\beta}_k} ).
\end{equation}
%
%Thus $\forall k$, for $\hat{\boldsymbol{m}}_{\cdot k}$'s value, we get:
%\begin{equation}
%\hat{\boldsymbol{m}}_{\cdot k}=\sum_{i,t} \boldsymbol{m}_{ik}^t-\omega_{ik}^t
%\end{equation}

%According to \cite{antoniak1974mixtures, teh2006hierarchical, fox2008hdp}, $\boldsymbol{m}$'s value is distributed in proportional to $s(n_{ik}^t, m)(\alpha\boldsymbol{\beta}_k)^m$ (here $s(n,m)$ are unsigned Stirling numbers of the first kind). In our sticky case, part of the label are generated as the previous ones, which is not our desired. Thus, we need to generate variables $\omega_{ik}^t$
\end{description}

\subsection{Adapted Slice-Efficient Sampling}

We also incorporate the slice-efficient sampling \cite{Kalli:2011:SSM:1900672.1900678}\cite{walker2007sampling} to our model. The original sampling scheme was designed to sample the Dirichlet Process Mixture model. In order to adapt it to our framework, which is based on a HDP prior and also has pair-wise membership indicators, we use auxiliary variables $U = \{u_{ij,s}^t, u_{ij, r}^t\}$ for each of the latent membership pair $\{s_{ij}^t, r_{ij}^t\}$. Having the $U$s, we are able to limit the number of components in which ${\boldsymbol{\pi}_i}$ needs to be considered, which is infinite otherwise.

Under the slice-efficient sampling framework, the variables of interest are now extended to: $\boldsymbol{\pi}_i^t, \{u_{ij,r}^t, u_{ij,s}^t\}, \{s_{ij}^t, r_{ij}^t\}, \boldsymbol{\beta}, \boldsymbol{m}$:
\begin{description}

\item{Sampling $\boldsymbol{\pi}$}

For each node $i=1, \cdots, N$: we generate $\boldsymbol{\pi}_i^{'t}$ using sticky-breaking process \cite{Ishwaran01gibbssampling}, where each $k^{\text{th}}$ component is generated using:

$\boldsymbol{\pi}_{ik}^{'t}\sim $beta$(\boldsymbol{\pi}_{ik}^{'t};a_{ik}^t, b_{ik}^t)$, where
\begin{equation} \label{eq2}
\begin{split}
& a_{ik}^t=\alpha\boldsymbol{\beta}_k+N_{ik}^t+\kappa N_{ik}^{t-1}\\
& b_{ik}^t=\alpha(1-\sum_{l=1}^{k}\boldsymbol{\beta}_l)+N_{i,k_0>k}^t+\kappa N_{i,k^0>k}^{t-1}
\end{split}
\end{equation}
Here ${\boldsymbol{\pi}}_k={\boldsymbol{\pi}}_k^{'}\prod_{i=1}^{k-1} (1-{\boldsymbol{\pi}}_i^{'})$.

\item{Sampling $u_{ij,s}^t, u_{ij, r}^t, s_{ij}^t, r_{ij}^t$}

We use $u_{ij,s}^t\sim U(0, \boldsymbol{\pi}_{is_{ij}^t}^t)$, $u_{ij,r}^t\sim U(0, \boldsymbol{\pi}_{jr_{ij}^t}^t)$. Then the obtained hidden label is independently sampled from the finite candidates:
\begin{equation} \label{eq_1}
\begin{split}
&P(s_{ij}^t=k, r_{ij}^t=l|Z,e_{ij}^t, \boldsymbol{\beta}, \alpha, \kappa, N, \boldsymbol{\pi}, u_{ij,s}^t, u_{ij,r}^t))\\
 \propto & \boldsymbol{1}(k:\boldsymbol{\pi}_{ik}^t>u_{ij,s}^t)\cdot \boldsymbol{1}(l:\boldsymbol{\pi}_{jl}^t>u_{ij,r}^t) \\
& \cdot\prod_{l=1}^{2n}P(z_{il}^{t+1}|z_{i\cdot}^t\slash s_{ij}^t,s_{ij}^t=k, \boldsymbol{\beta}, \alpha, \kappa, N_i^{t+1})\\
&\cdot\prod_{l=1}^{2n}P(z_{jl}^{t+1}|z_{j\cdot}^t\slash r_{ij}^t,r_{ij}^t=l, \boldsymbol{\beta}, \alpha, \kappa, N_j^{t+1})\\
\cdot & P(e_{ij}^t|E\backslash\{e_{ij}^t\}, s_{ij}^t=k, r_{ij}^t=l, \boldsymbol{Z}\backslash\{s_{ij}^t, r_{ij}^t\}, \lambda_1, \lambda_2)\\
\end{split}
\end{equation}

\item{Sampling $\boldsymbol{\beta}$, $\boldsymbol{m}$}

This is the same as the Gibbs sampling.

\end{description}

The derivations of the above equations can be referenced to the Supplementary Materials.

\subsection{Hyper-parameter Sampling}
The hyper-parameters involved are ${\gamma},\alpha,\kappa$. However, it is impossible to compute their posterior individually. Therefore, we place three prior distributions on some ``combination'' of the variables: A vague gamma prior $\mathcal{G}(1, 1)$ is placed on both ${\gamma}, (\alpha+\kappa)$. A beta prior is placed on the ratio $\frac{\kappa}{\alpha+\kappa}$.

To sample $\gamma$ value, since $\log(\gamma)$'s posterior distribution is log-concave, we use the Adaptive Rejection Sampling (ARS) method \cite{rasmussen2000infinite}.

To sample $(\alpha+\kappa)$, we use the Auxiliary Variable Sampling \cite{teh2006hierarchical} using the auxiliary variable $\boldsymbol{m}$ in Eq. (\ref{eq_3}) as proposed in \cite{teh2006hierarchical}.

To sample $\frac{\kappa}{\alpha+\kappa}$, we place a vague beta prior $\mathcal{B}(1, 1)$ on it, with a likelihood of $\{\boldsymbol{m}_{ik}^t-\hat{\boldsymbol{m}}_{ik}^t, \forall i,k,t>1\}$ in Eq. (\ref{eq_4}), the posterior is in an analytical and samplable form, thanks to its conjugate property.

\subsection{Discussions}
Both the Gibbs Sampling and Slice-Efficient Sampling are two feasible ways of accomplishing our task. They have different pros and cons.

%favours as their intrinsic different sampling paradigm.

As mentioned previously, Gibbs Sampling in our \emph{DIM3} integrates out the mixed-membership distribution $\{\boldsymbol{\pi}_i^t\}$. It is the ``marginal approach'' \cite{papaspiliopoulos2008retrospective}. The property of community exchangeability makes it simple to implement. However, theoretically, the obtained samples mix slowly as the sampling of each label is dependent on other labels.

%and large cluster shift may occur in one iteration.

The Slice-Efficient Sampling is one ``conditional approach'' \cite{Kalli:2011:SSM:1900672.1900678} while the membership indicators are independently sampled from $\{\boldsymbol{\pi}_i^t\}$. In each iteration, given $\{\boldsymbol{\pi}_i^t\}$, we can parallelize the process of sampling membership indicators, which may help to improve the computation, especially when the number of nodes, i.e., $N$ becomes larger, and the number of communities, i.e., $k$ becomes smaller.

\begin{figure*}[!ht]
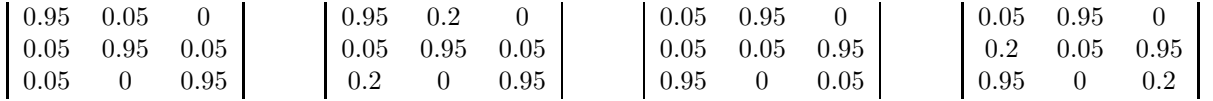

  \begin{minipage}[b]{0.24\textwidth}
    \centering
       \begin{tabular}{|ccc|}
    0.95 & 0.05 & 0  \\
    0.05 & 0.95 & 0.05 \\
    0.05 & 0 & 0.95 \\
       \end{tabular}
  \end{minipage}
  \begin{minipage}[b]{0.24\textwidth}
    \centering
       \begin{tabular}{|ccc|}
    0.95 & 0.2 & 0  \\
    0.05 & 0.95 & 0.05 \\
    0.2 & 0 & 0.95 \\
        \end{tabular}
  \end{minipage}
      \begin{minipage}[b]{0.24\textwidth}
    \centering
       \begin{tabular}{|ccc|}
    0.05 & 0.95 & 0  \\
    0.05 & 0.05 & 0.95 \\
    0.95 & 0 & 0.05 \\
       \end{tabular}
  \end{minipage}
    \begin{minipage}[b]{0.24\textwidth}
    \centering
       \begin{tabular}{|ccc|}
    0.05 & 0.95 & 0  \\
    0.2 & 0.05 & 0.95 \\
    0.95 & 0 & 0.2 \\
       \end{tabular}
  \end{minipage}
      \caption{Four Cases of the Compatibility Matrix. (Cases 1-4 as from left to right.)} \label{fig_2}
\end{figure*}

\section{Experiments} \label{sec_4}
The performance of the \emph{DIM3} model is validated by experiments on synthetic datasets and several real-world datasets. We implement the our model's finite-communities case as a baseline algorithm, namely as \emph{f-MTV} and \emph{f-MTI}.

\subsection{Synthetic Dataset} \label{sec_41}
For the synthetic data generation, the variables are generated following \cite{DBLP:journals/jmlr/HoSX11}. We use $N=20, T=3$, and hence $E$ is a $20\times20\times3$ asymmetric and binary matrix. The parameters are set up such that the $20$ nodes are equally partitioned into 4 groups. The ground-truth of the mixed-membership distribution for each of the groups are: $[0.8, 0.2, 0; 0, 0.8, 0.2; 0.1, 0.05, 0.85; 0.4, 0.4, 0.2]$.

%The synthetic dataset's generation follows the same procedure as \cite{DBLP:journals/jmlr/HoSX11}. We generate the synthetic dataset with $n=20$ nodes and $t=3$ time intervals. These

We consider 4 different cases to fully assess \emph{DIM3} against the ground-truth, all lie in the 3-role compatibility matrix.
\begin{description}
\item{{\bf Case 1}}: large diagonal values and small non-diagonal values
\item{{\bf Case 2}}: large diagonal values and mediate non-diagonal values
\item{{\bf Case 3}}: large non-diagonal values and small diagonal values
\item{{\bf Case 4}}: small diagonal values and mediate non-diagonal values
\end{description}

%The top row of Figure \ref{fig_3}.-\ref{fig_4}. is the manually 3-roles compatibility matrix of each case, while the bottom row presents the visualization of synthetic relational dataset at time point 1 and 2, with the black zone denotes 1 and while zone being 0.
The detailed value of the role-compatibility matrix on these four  cases are shown in Figure \ref{fig_2}.
%The top row of Figure \ref{fig_2}. expresses the manually setup role-compatibility matrix values. We choose these four cases as to explore our \emph{DIM3}'s recovery ability while different values of inter-community and intra-community connection may occur in real world.

\subsubsection{MCMC Analysis}
The convergence behavior is tested in terms of two quantities: the cluster number $K$, i.e., the number of different
values $Z$ can take, and the estimated density $D$ \cite{Kalli:2011:SSM:1900672.1900678, papaspiliopoulos2008retrospective}, which is defined as:
\begin{equation}
D = -2\sum_{i,j,t}\log\left(\sum_{k,l}\frac{N_{ik}^t\cdot N_{jl}^t}{4n^2T}p(e_{ij}^t|Z, \lambda_1, \lambda_2)\right)
\end{equation}

In our MCMC stationary analysis, we ran 5 independent Markov chains and discarded the first half of the Markov chains as a burn-in. With the random partition of 3 initial classes as the starting point, $130,000$ iterations are conducted in our samplings.
%
%\begin{table*}[ht]
%\caption{Synthetic data MCMC Convergence Diagnostics} \label{table_1}
%\centering
%\begin{tabular}{c|c|c|cc|c|c}
%  \hline
%  % after \\: \hline or \cline{col1-col2} \cline{col3-col4} ...
%  \multirow{2}{*}{Diagnostics} &   \multirow{2}{*}{Sampling}&   \multirow{2}{*}{variable} & \multicolumn{2}{c|}{G. \& R. D.} & Geweke D. & H. \& W. D. \\ \cline{4-7}
%   & & &PSRF & C.I. & $Z$-score & $p$-value\\
% \hline
%   \multirow{4}{*}{Case 1} & \multirow{2}{*}{Gibbs} & K  & 5.681 & 5.653 &	5.631 & 1 \\
%  &  & D & 0.754 &	0.662 & 0.586  &1 \\ \cline{2-7}
%  & \multirow{2}{*}{Slice}  & K & 0.754 &	0.662 & 0.586  &1 \\
%  & & D & 0.754 &	0.662 & 0.586  &1 \\
%  \hline
%   \multirow{4}{*}{Case 1} & \multirow{2}{*}{Gibbs} & K  & 5.681 & 5.653 &	5.631 & 1 \\
%  &  & D & 0.754 &	0.662 & 0.586  &1 \\\cline{2-7}
%  & \multirow{2}{*}{Slice}  & K & 0.754 &	0.662 & 0.586  &1 \\
%  & & D & 0.754 &	0.662 & 0.586  &1 \\
%  \hline
%   \multirow{4}{*}{Case 1} & \multirow{2}{*}{Gibbs} & K  & 5.681 & 5.653 &	5.631 & 1 \\
%  &  & D & 0.754 &	0.662 & 0.586  &1 \\\cline{2-7}
%  & \multirow{2}{*}{Slice}  & K & 0.754 &	0.662 & 0.586  &1 \\
%  & & D & 0.754 &	0.662 & 0.586  &1 \\
%  \hline
%   \multirow{4}{*}{Case 1} & \multirow{2}{*}{Gibbs} & K  & 5.681 & 5.653 &	5.631 & 1 \\
%  &  & D & 0.754 &	0.662 & 0.586  &1 \\\cline{2-7}
%  & \multirow{2}{*}{Slice}  & K & 0.754 &	0.662 & 0.586  &1 \\
%  & & D & 0.754 &	0.662 & 0.586  &1 \\
%  \hline
%\end{tabular}
%\end{table*}

The simulated chains satisfy standard convergence criteria, as we implemented the test by using CODA package \cite{coda_package}. In Gelman and Rubin's diagnostics \cite{gelman1992inference}, the value of Proportional Scale Reduction Factor (PSRF) is 1.09 (with upper C.I. 1.27) for k, 1.03 (with upper C.I. 1.09) for D in the Gibbs sampling, and 1.02 (with upper C.I. 1.06) for k, 1.02 (with upper C.I. 1.02) for D in Slice sampling. The Geweke's convergence diagnostics \cite{Geweke92evaluatingthe} is also employed, with proportion of first $10\%$ and last $50\%$ of the chain as comparison. The corresponding z-scores are all in the interval $[-2.09, 0.85]$ for 5 chains. In addition, the stationarity and half-width tests of Heidelberg and Welch Diagnostic \cite{Heidelberger:1981:SMC:358598.358630} were both passed in all the cases, with $p$-value higher than 0.05. Based on all these statistics, the Markov chain's stationary can be safely ensured in our case.

The efficiency of the algorithms can be measured by estimating the integrated autocorrelation time $\tau$ for $K$ and $D$. $\tau$ is a good performance indicator as it measures the statistical error of Monte Carlo approximation on a target function $f$. The smaller $\tau$, the more efficient of the algorithm.

\cite{Kalli:2011:SSM:1900672.1900678} used an estimator $\widehat{\tau}$ as:
\begin{equation}
\widehat{\tau}=\frac{1}{2}+\sum_{l=1}^{C-1}\widehat{\rho}_l
\end{equation}
Here $\widehat{\rho}_l$ is the estimated autocorrelation at lag $l$ and $C$ is a cut-off point, which is defined as $C:=\min\{l:|\widehat{\rho}_l|<2/\sqrt{M}\}$, and $M$ is the number of iterations.

\begin{table*}[!ht]
\caption{Integrated Autocorrelation Times Estimator $\widehat{\tau}$ for $K$ and $D$} \label{table_2}
\centering
\begin{tabular}{c|c|c|c|c|c|c|c|c|c|c|c}
  \hline
  \multicolumn{2}{c|}{} &\multicolumn{5}{c|}{K}&\multicolumn{5}{c}{D}\\
  \hline
Sampling & \backslashbox{$\gamma$}{$\alpha$} & 0.1 & 0.3 & 0.5 & 1 & 2&0.1 & 0.3 & 0.5 & 1 & 2 \\
  \hline
\multirow{5}{*}{\emph{MTV-g}}& 0.1 & 177.2  & 93.65 & 26.91 & 50.21 & 11.24 & 358.8 & 148.3 & 23.94 & 84.75 & 4.31\\ \cline{2-12}
& 0.3 & 260.5 & 54.00 & 9.18 & 5.31 & 6.56 & 389.5 & 315.0 & 3.11 & 26.32 & 4.78\\ \cline{2-12}
& 0.5 & 1.83  & 8.33 & 7.54 & 3.95 & 5.24 & 2.88 & 79.34 & 90.93 & 3.17 & 3.82\\ \cline{2-12}
& 1.0 & 5.57 & 6.45 & 3.44 & 3.64 & 4.56 & 3.19 & 2.78 & 1.76 & 8.14 & 5.74\\ \cline{2-12}
& 2.0 & 4.30  & 2.87 & 3.35 & 2.98 & 3.28 & 95.48 & 1.91 & 3.29 & 8.74 & 6.55 \\
\hline
\hline
\multirow{5}{*}{\emph{MTV-s}}& 0.1 & 248.6 & 90.63 & 161.3 & 9.58 & 17.69 &8.67 & 59.90 & 57.57 & 1.87 & 3.70\\ \cline{2-12}
&0.3 & 120.6 & 66.23 & 44.35 &  11.40 & 7.28 & 29.05 & 20.64 & 30.01 & 45.57 & 3.40\\ \cline{2-12}
& 0.5 & 18.99 & 27.27 & 6.08 & 8.76 & 10.40 &39.66 & 3.87 & 5.30 & 3.17 & 5.83\\ \cline{2-12}
& 1.0 & 5.79 & 9.19 & 11.85 & 8.46 & 7.25 & 40.51 & 4.85 & 3.12 & 6.88 & 10.51\\ \cline{2-12}
& 2.0 & 3.17 & 8.41 & 5.35 & 5.48 & 5.05 & 25.54 & 34.82 & 4.61 & 35.61 & 12.68\\
  \hline
\end{tabular}
\end{table*}

We test the sampling efficiency of \emph{MTV-g} and \emph{MTV-s} on Case 1 with the same setting as \cite{papaspiliopoulos2008retrospective}. Among the whole $130,000$ iterations, the first $30,000$ samples are discarded as a burn-in and the rest is thinned $1/20$. We manually try different values of the hyper-parameters $\gamma$ and $\alpha$ and show the integrated autocorrelation time estimator in Table \ref{table_2}. Although some outliers exist, we can see that there is a general trend that, with fixed $\alpha$ value, the autocorrelation function will decrease when the $\gamma$ value increases. This same phenomenon happens on $\alpha$ while $\gamma$ fixed. This fact meets our empirical knowledge. The larger value of $\gamma, \alpha$ will help to discover more clusters, then comes a smaller autocorrelation function.

On the other hand, we admit that \emph{MTV-g} and \emph{MTV-s} do not show much difference in the Markov chain's mixing rate as shown in Table \ref{table_2}. As mentioned in the previous section, Slice sampling provides an mixed-membership distribution independent sampling scheme, which can enjoy the time efficiency of parallel computing in one iteration. For large scale datasets, it is an feasible solution. While in Gibbs sampling, the parallel computing is impossible as the sampling variables are in a dependent sequence.

%
%Since we are always seeking a faster mixing markov chain, i.e., less correlation between the obtained samples, the smaller value of the autocorrelation value, the larger efficient sample size, the better. The average effective sample size is $92.9$ per 10,000 samples in gibbs sampling and $65.4$ in slice sampling. Thus, the gibbs sampling mixes faster than the slice sampling. In our conjecture, this is due to the instant influence of labels changing in gibbs sampling, while in slice sampling the changing label's influence will be slow as one iteration finishes. However, since the label sampling are all independent variables given the mixed-membership distribution $\boldsymbol{\pi}_i^t$, this could be achieved by using the parallel computing. While in gibbs sampling, as it is a sequential sampling, the process could not be further improved.

\subsubsection{Further Performance}

\begin{figure}[htbp]
\centering
\includegraphics[scale=0.5, width = 0.5 \textwidth]{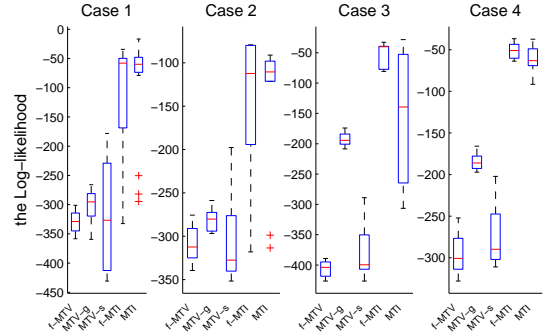}
\caption{Log-likelihood Performance}
\label{fig_6}
\end{figure}

\begin{table*}[htbp]
\caption{Average $l_2$ Distance to the Ground-truth} \label{table_1}
\centering
\begin{tabular}{c|c|c|c|c|c|c|c|c|c|c}
  \hline
\multirow{2}{*}{Cases} &\multicolumn{5}{c|}{Role-Compatibility Matrix}&\multicolumn{5}{c}{Mixed-Memberships}\\
  \cline{2-11}
  &\emph{f-MTV} & \emph{MTV-g} & \emph{MTV-s} & \emph{f-MTI} & \emph{MTI}& \emph{f-MTV} & \emph{MTV-g} & \emph{MTV-s} & \emph{f-MTI} & \emph{MTI} \\
  \hline
1 & $0.529$ & $0.625$ & $0.848$ & $0.114$ & $\boldsymbol{0.086}$ & $0.366$ & $0.384$ & $0.403$ & $0.199$ & $\boldsymbol{0.191}$ \\
\hline
2 & $0.439$ & $0.225$ & $0.339$ & $\boldsymbol{0.195}$& $0.204$ & $0.355$ & $0.355$ & $0.319$ &$\boldsymbol{0.207}$ &  $0.227$ \\
\hline
3 & $0.134$ & $0.201$ & $0.513$ &$0.117$ & $\boldsymbol{0.087}$ & $0.278$ & $0.289$ & $0.589$ & $0.208$ & $\boldsymbol{0.187}$ \\
\hline
4 & $\boldsymbol{0.195}$ & $0.214$ & $0.267$& $0.220$ & $0.219$ & $0.258$ & $0.285$ & $0.277$ & $0.192$ & $\boldsymbol{0.182}$ \\
\hline
\end{tabular}
\end{table*}

We will compare the models in terms of the Log-likelihood (in Figure \ref{fig_6}); the average $l_2$ distance between the mixed-membership distributions and its ground-truth; and the one between the posterior role-compatibility matrix and its ground-truth (in Table \ref{table_1}).

%Both the \emph{MAP} and \emph{MLE} estimators are employed to determine the best performed sample. However, in our empirical finding, we find \emph{MLE} always works better. Thus, we choose maximized marginal log-likelihood sample (i.e. \emph{MLE}) as our preferred result. Among the multiple results we obtained from the data, we choose the one with the best performance and present it in Table \ref{table_1}.

From the log-likelihood comparison in Figure \ref{fig_6}, we can see that the \emph{MTI} model performs better than the \emph{MTV} model. On the average $l_2$ distance to the ground-truth performance, the \emph{MTI} model also performs better.
%The results exhibit in Figure \ref{fig_2}. We can see that our model can in generally recover the role-compatibility matrix on all the four cases. The large compatibility value are still kept, and even with the mediate value (0.2), our model can still distinct it from the small value(0.05) in general.

%Our performance is held to compare with , in terms of mixed-membership distribution recovery rate $err(\pi)$ and role-compatibility matrix recovery variance $var(W)$. With the true community number setting as their parameter, we run each model 200 times and select one with the highest BIC score.

%Table \ref{table_1}. represents our performance on the dynamic mixed-membership distribution Average $l_2$ distance to the ground-truth mixed-membership in section \ref{sec_41}. Also we provide the performance on the \emph{dMMSB} \cite{xing2010state}.
%
%
%The Average $l_2$ distance of our method is at least as small as the \emph{dMMSB}. Considering we do not require the community number as prior information, our method is more adaptable.

Here we compare the computational complexity (Running Time) of the models in one iteration, with $K$ discovered communities and show the results in Table \ref{table_3}. We discuss \emph{MTV-g} and \emph{MTV-s} as an instance. In \emph{MTV-g}, the number of variables to be sampled is $(2K+2n^2T)$ , while a total of $(2K+4n^2T+nT)$ variables are sampled in \emph{MTV-s}.
However, the posterior calculation of $Z$ in \emph{MTV-s} can be directly obtained from the membership distribution, while we need to calculate the ratio for each of $Z$ in \emph{MTV-g}. Also, the $U$ value at each time can be sampled in one operation as its independency in \emph{MTV-s}. Thus, the result of \emph{MTV-s} runs faster than \emph{MTV-g} is in accordance with our assumption.
\begin{table}[!tp]
\caption{Running Time (Seconds per iteration)} \label{table_3}
\centering
\begin{tabular}{c|c|c|c|c|c}
  \hline
N.  &\emph{f-MTV} & \emph{MTV-g} & \emph{MTV-s} & \emph{f-MTI} & \emph{MTI}  \\
  \hline
20  &$0.20$ & $0.28$ & $0.23$ & $ 0.15$ & $ 0.31$ \\
\hline
50  & $ 1.03$  & $ 1.52$ & $ 1.29$ & $0.95$ &$1.91$ \\
\hline
100 & $3.69$ & $5.76$ & $4.81$ & $3.74$ & $7.49$ \\
\hline
200 & $15.61$ & $24.17$ & $19.87$ & $15.82$ & $30.19$ \\
\hline
500 & $ 106.96$ & $ 154.45$ & $ 119.82$ & $105.61$ & $202.09$ \\
\hline
1000 & $493.44$ & $ 888.86$ & $ 642.28$ & $597.29$ & $1102.90$ \\
\hline
\end{tabular}
\end{table}

\subsection{Real World Dataset Performance}
We randomly selected 7 real world datasets for benchmark testing. Their detailed information, including the number of nodes, the number of edges, edge types and time intervals, are give in Table \ref{table_4}. We will discuss more on the first two datasets in the following.

\subsubsection{Log-likelihood Performance on Various Real World Datasets}

\begin{table}[htbp]
\caption{Data Set Information} \label{table_4}
\centering
\begin{tabular}{c|c|c|c|c}
  \hline
Dataset & Nodes & Edge & Time & Type\\
  \hline
  Kapferer & 39 & 256 & 2 & friends \\
  \hline
    Sampson & 18 & 168 & 3 & like \\
  \hline
  Stu-net & 50 & 351 & 3 & friends \\
  \hline
  Enron & 41 & 1980 & 12 & email \\
  \hline
  Newcomb & 17 & 1020 & 15 & contact \\
  \hline
  Freeman & 32 & 357 & 2 &  friends\\
  \hline
  Coleman & 73 & 506 & 2 & co-work \\
  \hline
  \end{tabular}
\end{table}

Since no unambiguous ground-truth can be found in the real world dataset, we mainly use log-likelihood to verify the corresponding model's performance: The larger the log-likelihood, the better appropriateness of the model to data.

\begin{table*}[!tp]
\caption{Log-likelihood Performance ($95\%$ Confidence Interval = Mean $\mp1.96*$Standard Error)} \label{table_3}
\centering
\begin{tabular}{c||c|c|c|c|c}
  \hline
Dataset & \emph{f-MTV} & \emph{MTV-g} & \emph{MTV-s} & \emph{f-MTI} &\emph{MTI} \\
  \hline
  Kapferer  & $-247.4 \mp 28.9$ &$-267.7 \mp 36.3$ &$-332.6 \mp 51.3$ & $\boldsymbol{-43.4 \mp 0.5}$& $-88.9 \mp 4.4$\\
    \hline
  Sampson  & $-290.0 \mp 59.4$ &$-219.2 \mp 8.4$ &$-256.4 \mp 11.1$ & $-79.3 \mp 5.9$& $\boldsymbol{-53.3 \mp 4.1}$\\
  \hline
  Stu-net  & $-574.5 \mp 18.4$ &$-505.8 \mp 32.1$ &$-506.3 \mp 21.2$ & $\boldsymbol{-42.3 \mp 18.0}$ & ${-47.2 \mp 7.1}$\\
  \hline
  Enron  & $-2398.3 \mp 75.4$& $-2701.4 \mp 51.3$& $-2489.9 \mp 55.4$& $\boldsymbol{-656.8 \mp 56.4}$&$-1368.3 \mp 23.1$\\
  \hline
  Newcomb  & $-1342.7 \mp 32.1$ &$-1294.1 \mp 41.7$ &$-1320.6 \mp 32.9$ &$-444.0 \mp 26.3$ & $\boldsymbol{-343.4 \mp 34.5}$\\
  \hline
  Freeman  & $-378.9 \mp 28.6$ &$-406.8 \mp 37.8$ &$-406.9 \mp 49.5$ & $\boldsymbol{-22.5 \mp 3.1}$ & ${-27.2 \mp 9.2}$\\
  \hline
  Coleman  & $-1321.8 \mp 116.7$ &$-1270.2 \mp 101.1$ & $-1283.9 \mp 168.3$ & $-54.8 \mp 39.3$ & $\boldsymbol{-41.9 \mp 1.0}$ \\
  \hline
  \end{tabular}
\end{table*}

Table \ref{table_3} shows the $95\%$ confidence interval in test data log-likelihood of our models versus the classical ones. The black bold type denotes the largest value in each of the rows. We can see the \emph{MTI} model usually performs better, while the \emph{MTV} model may be bothered by the over-fitting problem in our assumption.

\subsubsection{Kapferer Tailor Shop}
The Kapferer Tailor Shop data \cite{nowicki2001estimation} records interactions in a tailor shop at two time points. In this time period, the employees in the shop are negotiating for higher wages. The data set is of particular interesting as two strikes happen after each time point, with the first fails and the second succeeds.

We mainly use the ``work-assistance'' interaction matrix in the dataset. The employees have 8 occupations: head tailor (19), cutter (16), line 1 tailor (1-3, 5-7, 9, 11-14, 21, 24), button machiner (25-26), line 3 tailor (8, 15, 20, 22-23, 27-28), Ironer (29, 33, 39), cotton boy (30-32, 34-38) and line 2 tailor (4, 10, 17-18).

\begin{figure}[!tp]
\centering
\includegraphics[scale=0.4, width = 0.4 \textwidth]{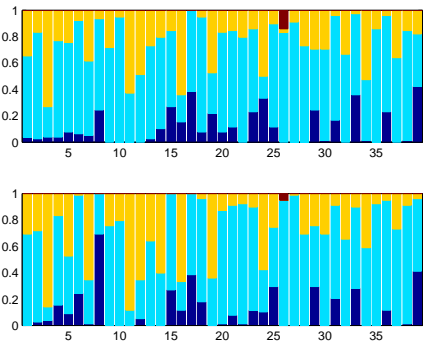}
\caption{\emph{MTI} Performance on Sampson Monastery Dataset (Top: Time 1; Bottom: Time 2.)}
\label{fig_9}
\end{figure}

%In more details, we find all the line 2-3 tailors belong to community B, this may be due to the working place restriction.

We can see the yellow bar at time point 2 are larger than the ones at time point 1, which means people tending to have another group at time point 2, rather than mostly dominated by one large groups at time point 1.

\subsection{Sampson Monastery Dataset}
The Sampson Monastery dataset are used here to do an exploratory study. There are 18 monks in this dataset, and their social linkage data is collected at 3 different time points with various relations. Here we especially focus on the like-specification. In the like-specification data, each monk selects three monks as his top-closed friends. In our settings, we mark the selected relations as 1, otherwise 0. Thus, an $18\times 18\times 3$ social networks data set is constructed, with each row has three elements valued 1.

According to the previous studies \cite{kim2012nonparametric, xing2010state}, the monks are divided into 4 communities: \emph{Young Turks}, \emph{Loyal Opposition}, \emph{Outcasts} and an interstitial group.
\begin{figure}[!tp]
\centering
\includegraphics[scale=0.4, width = 0.4 \textwidth]{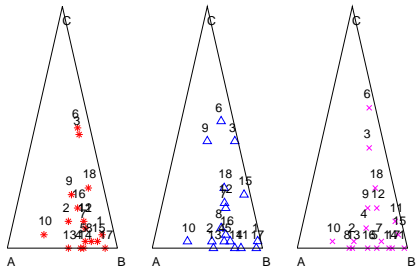}
\caption{\emph{MTI} Performance on Sampson Monastery Dataset (from Left to Right: Time 1-3.)}
\label{fig_1}
\end{figure}

Figure \ref{fig_1} shows the detailed results of \emph{MTI}. As three communities have been detected, we put all the results in a $2$-simplex, with which we denote as $A$, $B$ and $C$. As we can see, most of the monks stay the in the same area during the 3 time points, except for the monk $8$ and $9$.

We also provide the role-compatibility matrix in Figure \ref{fig_7} for comparison. Compared to the result in \cite{xing2010state}, our results are with a larger compatibility value within the same role. Also, the first role's value in our model is 0 while it is about 0.6 in \cite{xing2010state}.
    \begin{figure}[!ht]
    \centering
  \begin{minipage}[b]{0.19\textwidth}
    \centering
       \begin{tabular}{|ccc|}
      0.09 & 0 & 0.0  \\
    0.05 & 0.99 & 0.02 \\
    0.01 & 0 & 0.96 \\
       \end{tabular}
  \end{minipage}
  \begin{minipage}[b]{0.19\textwidth}
    \centering
       \begin{tabular}{|ccc|}
      0.01 & 0 & 0.03  \\
    0.02 & 0.78 & 0 \\
    0.02 & 0 & 0.67 \\
        \end{tabular}
  \end{minipage}
  \caption{Role Compatibility Matrix (Left: \emph{MTV-g}; Right: \emph{MTI})} \label{fig_7}
\end{figure}

\section{Conclusion \& Future Work} \label{sec_5}
In this paper, we have extended the existed mixed-membership stochastic blockmodel to the infinite community case in the dynamic setting. By incorporating the mixed-membership distribution-sticky paradigm, we have realized the time-correlation description on the hidden label. Both the Gibbs sampling and adapted Slice-Efficient sampling have been utilized to achieve the inference target. Quantity analysis on the MCMC's convergence behaviour, including the convergence test, autocorrelation function, etc., have been provided to further enhance the inference performance. The results in the experiments verify that our \emph{DIM3} is effective to re-construct the dynamic mixed-membership distribution and the role-compatibility matrix.

Some future work includes a systematic application of DIM3 to various large real-world social networks. In particular, we are also interested in adapting our model to many atypical applications, for example, where sequences of networks have non-binary and directional measurements. We will also study other more flexible framework for modelling persistence of memberships over times. Lastly, we will perform an extensive study into patterns of joint dynamics of $\{\boldsymbol{\pi}_i^t\}$ and to extract meaningful latent information from them. This is done in a setting where the number of components between $\boldsymbol{\pi}_i^{t_1}$ and $\boldsymbol{\pi}_i^{t_2}$ may differ.

\bibliography{dnmdr}

\begin{thebibliography}{10}

\bibitem{airoldi2008mixed}
E.M. Airoldi, D.M. Blei, S.E. Fienberg, and E.P. Xing.
\newblock Mixed membership stochastic blockmodels.
\newblock {\em The Journal of Machine Learning Research}, 9:1981--2014, 2008.

\bibitem{fox2008hdp}
E.B. Fox, E.B. Sudderth, M.I. Jordan, and A.S. Willsky.
\newblock An hdp-hmm for systems with state persistence.
\newblock In {\em Proceedings of the 25th international conference on Machine
  learning}, pages 312--319. ACM, 2008.

\bibitem{fox2011bayesian}
Emily Fox, Erik~B Sudderth, Michael~I Jordan, and Alan~S Willsky.
\newblock Bayesian nonparametric inference of switching dynamic linear models.
\newblock {\em Signal Processing, IEEE Transactions on}, 59(4):1569--1585,
  2011.

\bibitem{fox2011sticky}
Emily~B Fox, Erik~B Sudderth, Michael~I Jordan, and Alan~S Willsky.
\newblock A sticky hdp-hmm with application to speaker diarization.
\newblock {\em The Annals of Applied Statistics}, 5(2A):1020--1056, 2011.

\bibitem{fu2009dynamic}
W.~Fu, L.~Song, and E.P. Xing.
\newblock Dynamic mixed membership blockmodel for evolving networks.
\newblock In {\em Proceedings of the 26th Annual International Conference on
  Machine Learning}, pages 329--336. ACM, 2009.

\bibitem{gelman1992inference}
A.~Gelman and D.B. Rubin.
\newblock Inference from iterative simulation using multiple sequences.
\newblock {\em Statistical science}, 7(4):457--472, 1992.

\bibitem{Geweke92evaluatingthe}
J.~Geweke.
\newblock Evaluating the accuracy of sampling-based approaches to the
  calculation of posterior moments.
\newblock In {\em Bayesian Statistics}, pages 169--193. University Press, 1992.

\bibitem{Heidelberger:1981:SMC:358598.358630}
Philip Heidelberger and Peter~D. Welch.
\newblock A spectral method for confidence interval generation and run length
  control in simulations.
\newblock {\em Commun. ACM}, 24(4):233--245, April 1981.

\bibitem{DBLP:journals/jmlr/HoSX11}
Qirong Ho, Le~Song, and Eric~P. Xing.
\newblock Evolving cluster mixed-membership blockmodel for time-evolving
  networks.
\newblock {\em Journal of Machine Learning Research - Proceedings Track},
  15:342--350, 2011.

\bibitem{conf/nips/IshiguroIUT10}
Katsuhiko Ishiguro, Tomoharu Iwata, Naonori Ueda, and Joshua~B. Tenenbaum.
\newblock Dynamic infinite relational model for time-varying relational data
  analysis.
\newblock In {\em NIPS}, pages 919--927. Curran Associates, Inc., 2010.

\bibitem{Ishwaran01gibbssampling}
Hemant Ishwaran and Lancelot~F. James.
\newblock Gibbs sampling methods for stick-breaking priors.
\newblock {\em Journal of the American Statistical Association}, 96:161--173,
  2001.

\bibitem{Kalli:2011:SSM:1900672.1900678}
Maria Kalli, Jim~E. Griffin, and Stephen~G. Walker.
\newblock Slice sampling mixture models.
\newblock {\em Statistics and Computing}, 21(1):93--105, January 2011.

\bibitem{kemp2006learning}
C.~Kemp, J.B. Tenenbaum, T.L. Griffiths, T.~Yamada, and N.~Ueda.
\newblock Learning systems of concepts with an infinite relational model.
\newblock In {\em Proceedings of the national conference on artificial
  intelligence}, volume~21, page 381. Menlo Park, CA; Cambridge, MA; London;
  AAAI Press; MIT Press; 1999, 2006.

\bibitem{kim2012nonparametric}
D.I. Kim, M.~Hughes, and E.~Sudderth.
\newblock The nonparametric metadata dependent relational model.
\newblock In {\em Proceedings of the 29th Annual International Conference on
  Machine Learning}. ACM, 2012.

\bibitem{koutsourelakis2008finding}
P.S. Koutsourelakis and T.~Eliassi-Rad.
\newblock Finding mixed-memberships in social networks.
\newblock In {\em Proceedings of the 2008 AAAI spring symposium on social
  information processing}, 2008.

\bibitem{nowicki2001estimation}
K.~Nowicki and T.A.B. Snijders.
\newblock Estimation and prediction for stochastic blockstructures.
\newblock {\em Journal of the American Statistical Association},
  96(455):1077--1087, 2001.

\bibitem{papaspiliopoulos2008retrospective}
O.~Papaspiliopoulos and G.O. Roberts.
\newblock Retrospective markov chain monte carlo methods for dirichlet process
  hierarchical models.
\newblock {\em Biometrika}, 95(1):169--186, 2008.

\bibitem{coda_package}
Martyn Plummer, Nicky Best, Kate Cowles, and Karen Vines.
\newblock Coda: Convergence diagnosis and output analysis for mcmc.
\newblock {\em R News}, 6(1):7--11, 2006.

\bibitem{rasmussen2000infinite}
Carl~Edward Rasmussen.
\newblock The infinite gaussian mixture model.
\newblock {\em Advances in neural information processing systems}, 12(5.2):2,
  2000.

\bibitem{teh2006hierarchical}
Y.W. Teh, M.I. Jordan, M.J. Beal, and D.M. Blei.
\newblock Hierarchical dirichlet processes.
\newblock {\em Journal of the American Statistical Association},
  101(476):1566--1581, 2006.

\bibitem{walker2007sampling}
S.G. Walker.
\newblock Sampling the dirichlet mixture model with slices.
\newblock {\em Communications in Statistics¡ªSimulation and
  Computation{\textregistered}}, 36(1):45--54, 2007.

\bibitem{xing2010state}
E.P. Xing, W.~Fu, and L.~Song.
\newblock A state-space mixed membership blockmodel for dynamic network
  tomography.
\newblock {\em The Annals of Applied Statistics}, 4(2):535--566, 2010.

\bibitem{Yangml11}
Tianbao Yang, Yun Chi, Shenghuo Zhu, Yihong Gong, and Rong Jin.
\newblock Detecting communities and their evolutions in dynamic social networks
  - a bayesian approach.
\newblock {\em Machine Learning}, 82(2):157--189, 2011.

\end{thebibliography}
\bibliographystyle{plain}

\end{document}